\documentclass[a4paper,11pt]{article}
\usepackage{jcappub} 
\usepackage{lineno}
\usepackage{orcidlink}

\arxivnumber{2510.26239} 
\title{A Large-Area Optical Time Projection Chamber for Hard X-ray Polarimetry with Directional Imaging of Low-Energy Electron Recoils}

\author[a,b,1]{Davide Fiorina\note{Corresponding author.}\orcidlink{0000-0002-7104-257X}}
\author[c]{Giorgio Dho\orcidlink{0000-0001-9454-9894}}
\author[a,b]{Elisabetta Baracchini\orcidlink{0000-0003-4686-128X}}
\author[d]{Paolo Soffitta\orcidlink{0000-0002-7781-4104}}
\author[a,b]{Samuele Torelli\orcidlink{0000-0003-3622-3524}}
\author[a,b]{David J. G. Marques\orcidlink{0000-0002-0013-6341}}
\author[d]{Enrico Costa}
\author[d]{Sergio Fabiani}
\author[d]{Fabio Muleri}
\author[c]{Giovanni Mazzitelli\orcidlink{0000-0003-2830-4359}}
\author[e,b]{Atul Prajapati\orcidlink{0000-0002-4620-440X}}

\affiliation[a]{Gran Sasso Science Institute,\\Viale F. Crispi, 7, 67100 L'Aquila, Italy}
\affiliation[b]{INFN Laboratori Nazionali del Gran Sasso,\\Via G. Acitelli, 22, 67100 Assergi (L'Aquila), Italy}
\affiliation[c]{INFN Laboratori Nazionali di Frascati,\\Via Enrico Fermi 54, 00044 Frascati (Roma), Italy}
\affiliation[d]{IAPS - INAF Istituto di Astrofisica e Planetologia Spaziali,\\Via del Fosso del Cavaliere 100, 00133 Roma, Italy}
\affiliation[e]{University of L'Aquila,\\Palazzo Camponeschi, Piazza Santa Margherita, 2, 67100 L'Aquila (AQ), Italy}

\emailAdd{davide.fiorina@gssi.it}

\abstract{We report on the development of a large-area, efficient, wide-field-of-view time projection chamber (TPC) for X-ray polarimetry, featuring a triple-GEM amplification stage and optical readout. When coupled with a light-weight capillary plate narrow-field collimator, it could measure polarization from bright galactic X-ray sources, while exposed to an open field, could detect polarization from transient objects on unexpected directions, such as Gamma-Ray Bursts. Originally developed within the CYGNO program for directional dark matter searches, the system employs a scientific CMOS (sCMOS) camera and a photomultiplier tube (PMT) to collect secondary scintillation light produced during charge amplification. A prototype with a cylindrical active volume (radius 3.7\,cm, height 5\,cm) was tested at the INAF--IAPS calibration facility (Rome, Tor Vergata) to assess sensitivity to low-energy electron directionality. We fully reconstruct electrons in the 10–60\,keV range, obtain angular resolutions as good as $15^\circ$, and infer modulation factors up to 0.9. These first results demonstrate robust photoelectron tracking at tens of keV with strong modulation, indicating that photoelectric-effect polarimetry can be extended to higher energies. This capability is promising for rapid transients (GRBs, solar flares) or bright ($>$ 100 mCrab) galactic sources, depending on the configuration of the collimator used, and would broaden the astrophysical reach of X-ray polarimetry.}

\begin{document}
\maketitle
\flushbottom

\section{Introduction} \label{sec:intro}
X-ray polarimetry has recently emerged as a powerful diagnostic in high-energy astrophysics, adding the degree and angle of polarization to the traditional observables (imaging, spectrum, timing) and enabling novel probes of magnetic-field geometry at the acceleration sites and radiation mechanisms and transport in compact objects.   Thanks to imaging, mapping of X-ray polarization is also now possible for extended sources, like Pulsar Wind Nebulae and SuperNova Remnants. \cite{Costa2001Nature, Weisskopf2010SPIE}. Advances in photoelectric polarimetry and the launch of NASA's Imaging X-ray Polarimetry Explorer (IXPE) \cite{10.1117/1.JATIS.8.2.026002,2021Soffitta} have made precise measurements routinely achievable, opening a new discovery space for extreme astrophysical environments. 
While imaging polarimetry requires an X-ray telescope, which can be demanding in terms of resources, non-imaging configurations with sufficient efficiency and large effective area could constitute the payload of a relatively small and fast X-ray mission, without the need for focusing optics.

When coupled with a wide-angle collimator, $\sim  40^\circ-50^\circ$  of aperture, photoelectric polarimetry may be employed for the study of Gamma-ray Bursts (GRBs) or magnetar bursts. 
At energies above $\sim 20~\mathrm{keV}$, several experiments based on Compton scattering (e.g., GAP, ASTROSAT, POLAR) have already been performed, although their results remain inconclusive and, in some cases, in tension with those obtained by \textit{INTEGRAL}. In the near future, \textit{POLAR-2} promises a significant improvement in sensitivity. However, the large effective area and the huge thickness required by Compton polarimeters generally leads to a large number of detector modules and increased system complexity.

When instead coupled to a narrow-field collimator ($\sim 1^\circ$--$2^\circ$), such a detector may provide rapid polarimetric monitoring of bright black-hole and neutron-star binaries. This observational configuration has already been demonstrated by PolarLight, a CubeSat polarimeter with an open geometrical area of only $1.6~\mathrm{cm}^2$ and a collimator aperture of $2.3^\circ$ operating in the 2--8~keV energy band. \textit{PolarLight} employed the same detector design as \textit{IXPE}, and measured a residual background of about $80$~mCrab. The background rejection efficiency of the \textit{IXPE} detector (approximately $40\%$) is known to be relatively small, primarily due to the trigger thresholds and gas gain settings adopted in its design \cite{Soffitta2012,DiMarco2023}. With Polarlight small effective area, the Crab Nebula, Sco X-1 and A0535+26 were monitored within the operative life of the experiment. 

Polarlight was only a smart pathfinder sharing space on a nanosat. But, by achieving significant results on a few very bright sources, it showed the path to future low mass, fast track experiments of tens or few hundreds square centimeter effective area.
The next generation of collimated experiments, both narrow- and wide-field, based on photoelectric effect, requires, especially at higher energies, detectors with a significantly larger collecting area. One possible approach is to tile multiple existing ASICs; for example, achieving a collection area of $100~\mathrm{cm}^{2}$ would require on the order of 40--50 ASICs, each with dimensions comparable to those used in the \textit{IXPE} detector. However, this solution is not optimal, due to the challenges associated with tiling and operating such a large number of ASICs, as well as the corresponding increase in power consumption. In addition, the effective area is reduced by the presence of dead regions around each ASIC to accommodate bonding pads and routing, and the need to ignore data from good pixels contigous to the dead regions wherever a uniform coverage of all azimuth angles would not be guaranteed.

The approach presented in this paper offers a way to overcome these limitations. It provides an efficient method to optically image tracks with the required spatial resolution (approximately $50~\mu\mathrm{m}$) over a collecting area $\ge 100~\mathrm{cm}^{2}$ using a single CMOS camera.

This approach has been explored in the past \cite{Austin1994,Sakurai2004}. The first demonstration achieved sensitivity above 40keV, while the second reached energies as low as 15 keV, obtaining a modulation factor of $\sim 25\%$ in an argon-based gas mixture at 1-atm. In both cases, a CCD camera was used to image the photoelectron tracks.

In this paper, we propose a novel implementation based on a triple-GEM amplification and scintillation stage, coupled to a high-efficiency lens and a state-of-the-art, ultra-low-noise CMOS camera. This implementation has been developed for directional dark-matter searches. This configuration enables high-resolution optical imaging of electron tracks with improved sensitivity and reduced background.

In the keV band, photoelectric polarimetry relies on reconstructing the emission angle of the photoelectron in the plane orthogonal to the incident photon direction. The same capability to image and reconstruct the initial direction of short electron tracks in gas is a defining feature of the CYGNO experiment, a directional dark matter program based on an optical time-projection chamber with triple-GEM amplification and optical readout \cite{amaro2022cygno}. In its primary scope, CYGNO targets the reconstruction of nuclear recoil directionality, while the same detector concept also supports a strong solar neutrino physics case through neutrino–electron elastic scattering, where keV scale electron recoils retain a correlation with the solar direction and provide an effective handle for background rejection and kinematic inference \cite{torelli2024feasibilitydirectionalsolarneutrino}. This focus on low-threshold electron tracking and angular reconstruction connects naturally to photoelectric polarimetry, where the polarization information is likewise encoded in the initial direction of an electron track. Building on the proven performance of the CYGNO 50 L prototype for X-ray photoelectron imaging \cite{commissionig} and on ongoing work on low energy electron recoil directionality in CYGNO \cite{directionality}, in this paper, we investigate the extension of this experimental approach to hard X-ray polarimetry with a wide field of view.

\subsection{Detector concept}
The proposed experimental approach consists of a gas-filled drift region operated at an electric field of order 1 kV/cm. Ionization electrons produced by a particle interacting in the gas drift toward a GEM-based \cite{Sauli:1997qp} amplification stage. In $CF_4$-rich gas mixtures, ionization and avalanche processes generate excited fragments such as $CF_3^\ast$ that de-excite via visible scintillation \cite{FRAGA200388}. 

A scientific CMOS (sCMOS) camera focused directly on the last GEM foil, coupled to a fast photographic lens with a large numerical aperture, is employed to efficiently collect scintillation light and to image the track 2D projection with sub-millimeter spatial resolution \cite{amaro2022cygno,commissionig}. A photomultiplier tube (PMT) records the time profile, enabling 3D track reconstruction and precise timing. Although this last feature is already routinely used in CYGNO, it will be applied in the context of celestial X-ray polarimetry with event rates much larger than those expected in the field of dark matter search; for this reason, this paper focuses only on 2D imaging, also for a direct comparison with the previous photoelectric polarimeters.

\begin{figure}[t!]
    \centering
    \begin{minipage}[b]{0.6\textwidth}
        \centering
        \includegraphics[width=\textwidth]{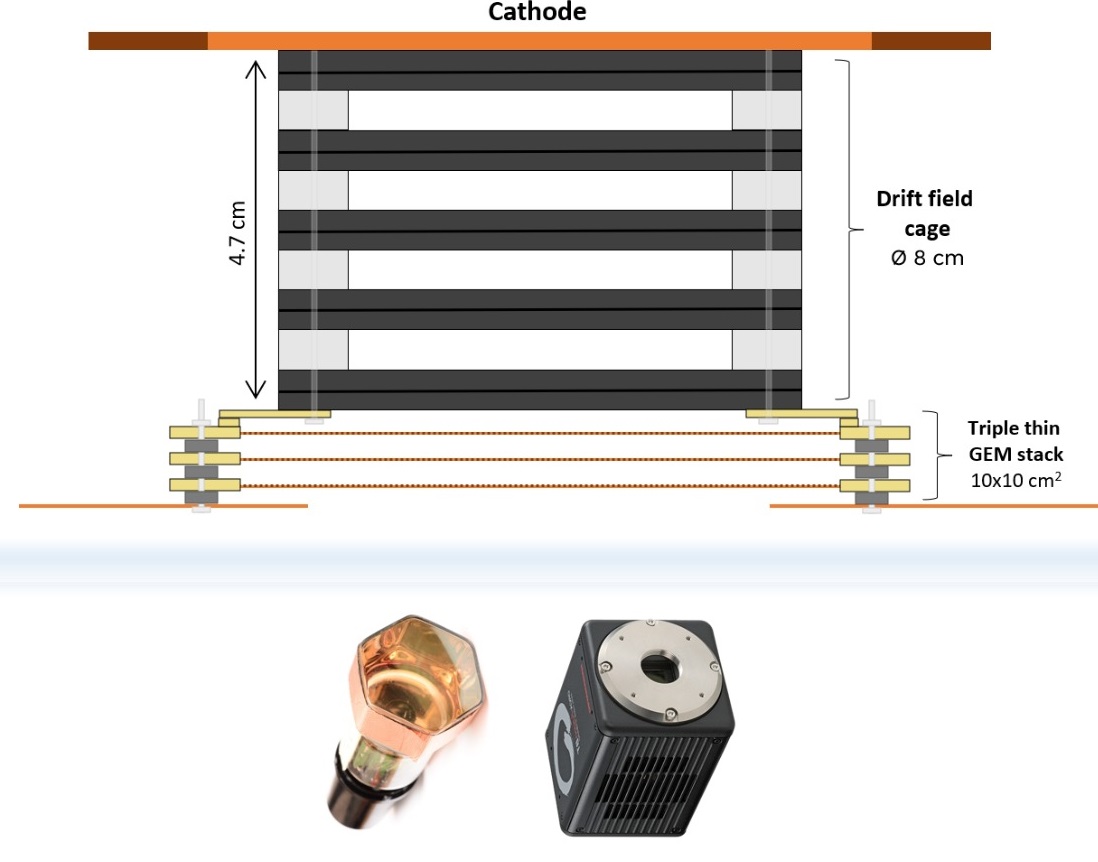}
        \caption{Schematic of the detector used in this work.}
        \label{fig:detSketch}
    \end{minipage}
    \hfill
    \begin{minipage}[b]{0.35\textwidth}
        \centering
        \includegraphics[width=\textwidth]{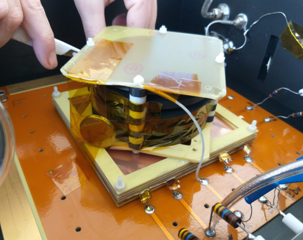}
        \caption{Photograph of the detector.}
        \label{fig:detPhoto}
        \vfill
        \includegraphics[width=\textwidth]{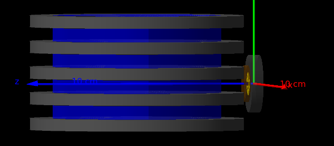}
        \caption{Geant4 simulation geometry for comparison with data.}
        \label{fig:detGeant}
    \end{minipage}
\end{figure}

The prototype detector used in this work features a standard 10 $\times$ 10 cm$^2$ thin triple-GEM amplification stage with 2 mm inter-GEM spacing operated with 2.5 kV/cm between them and the last electrode grounded. A cathode is positioned 5 cm above the first GEM, and a circular field cage of 7.4 cm diameter composed of silver wires encased in 3D printed plastic rings spaced about 1 cm ensures field uniformity in the drift region. The GEMs are powered by a custom floating High Voltage Power supply \cite{CORRADI200796}, a dedicated high voltage supply which works as an active high voltage (HV) divider. The TPC, typically operated in continuous gas flux mode, is enclosed in a 3D printed black plastic light-tight box that contains a gas-tight acrylic internal vessel. A thin window of highly transparent Mylar decouples the gas detector from the optical readout, which consists of a C14440-20UP ORCA-Fusion sCMOS camera placed at a distance of $(20.5 \pm 0.3)$ cm. A Schneider Xenon lens, coupled with the camera, focuses the last GEM plane. The camera possesses a squared silicon sensor of 1.498 $\times$ 1.498 cm$^2$ area, segmented in 2304 $\times$ 2304 pixels. Each pixel features a Quantum Efficiency (Q.E.) of about 80\% at 600 nm, and a readout noise of 0.7 electrons RMS. Within this scheme, the camera images an area of $\sim$ 11.3 $\times$ 11.3 cm$^2$, resulting in an effective pixel size of $\sim$ 49 $\times$ 49  $\mu$m$^2$. This pixel size is comparable to the pixel size of a Gas Pixel detector or of a GridPix detector, but on an area 60 times larger. Figures~\ref{fig:detSketch} and \ref{fig:detPhoto} show a schematic of the setup and actual picture of the detector, respectively, while Fig.~\ref{fig:detGeant} shows the relative Geant4 geometry simulated and used in the analysis discussed in Sec.\ref{sec:analysis}. For the measurements reported in this paper, not only the $ He/CF_4$60/40 gas mixture was employed representing the Dark Matter search standard in the CYGNO experiment context, but also alternative $He/CF_4$ ratios and $ Ar/CF_4$ mixtures were explored as a better choice for the polarimetric approach.

\subsection{Experimental Setup}
Before irradiating the detector with a polarized source, we first aimed to evaluate its capability to reconstruct the direction of tens-keV electrons in a gas mixture suitable for space-based operation. The final pressure, geometrical configuration, and overall design will be determined at a later stage through Monte Carlo simulations validated by comparison with experimental data. 
To generate electrons in the tens-of-keV range, we used a $\beta$-emitting radioactive source consisting of a ${}^{90}\mathrm{Sr}$ nuclide with an activity of $10~\mu\mathrm{Ci}$. This isotope undergoes $\beta$-decay, emitting electrons with a maximum energy of $546~\mathrm{keV}$. Its daughter nucleus, ${}^{90}\mathrm{Y}$, also undergoes $\beta$-decay, producing a second, higher-energy electron with a maximum energy of $2.28~\mathrm{MeV}$.
The $^{90}$Sr source is positioned between field-cage rings at the center of the drift gap, about 12 mm from the active volume, and a tungsten collimator (with 1 mm radius and 1 mm thickness) helps to select the emitted electron directions. Energy calibration is performed using  a $^{109}$Cd calibration source emetting X-rays of know energies (22 keV, 25 keV  and 88 keV) and using an $^{55}$Fe source (5.89 keV and 6.4 keV). We calibrated the prototype by fitting the integrated pixel counts along each reconstructed track as a function of the nominal absorbed X-ray energy. The Orca Fusion sCMOS images are collected by a custom DAQ software based on MIDAS in free running mode and with 0.3 s exposure for $^{90}$Sr and $^{55}$Fe and 0.22 s for $^{109}$Cd (due to the higher source activity).

\section{Low energy electrons detector performances}
In order to quantify the capability of the detector to measure the polarization of photons in the tens-of-keV energy range, we adopted the following procedure: 
(1) we reconstructed the direction of tens-of-keV photoelectrons using, as a proxy, the electron tracks produced by a ${}^{90}\mathrm{Sr}$ source, which are fully contained within the sensitive volume of the detector. This step requires deconvolution of the measured angular distribution from the intrinsic angular spread introduced by the collimator aperture, as evaluated with \textsc{Geant4} simulations \cite{geant}. From this, we derived the angular resolution as a function of electron energy; 
(2) we then folded this resolution, expressed as the rms of the distribution of reconstructed directions, with a $\cos^{2}\theta$ response to estimate the modulation factor $\mu$, defined as the detector response to a $100\%$ polarized beam; 
(3) finally, we combined $\mu$ with the intrinsic detection efficiency $\epsilon$, also evaluated by Monte Carlo simulation,  to calculate the figure of merit $\mu\sqrt{\epsilon}$, which directly enters the expression for the Minimum Detectable Polarization (MDP) \cite{figureof,Weisskopf2010SPIE,Kislat2015APh}. The MDP represents the minimum polarization fraction that can be detected from a celestial source at the $99\%$ confidence level (see par.\ref{sec:results}. for its definition).

\subsection{Data analysis}\label{sec:analysis}
\begin{figure}[t!]
    \centering
    \begin{minipage}[t]{0.3\textwidth}
        \centering
        \includegraphics[width=\textwidth]{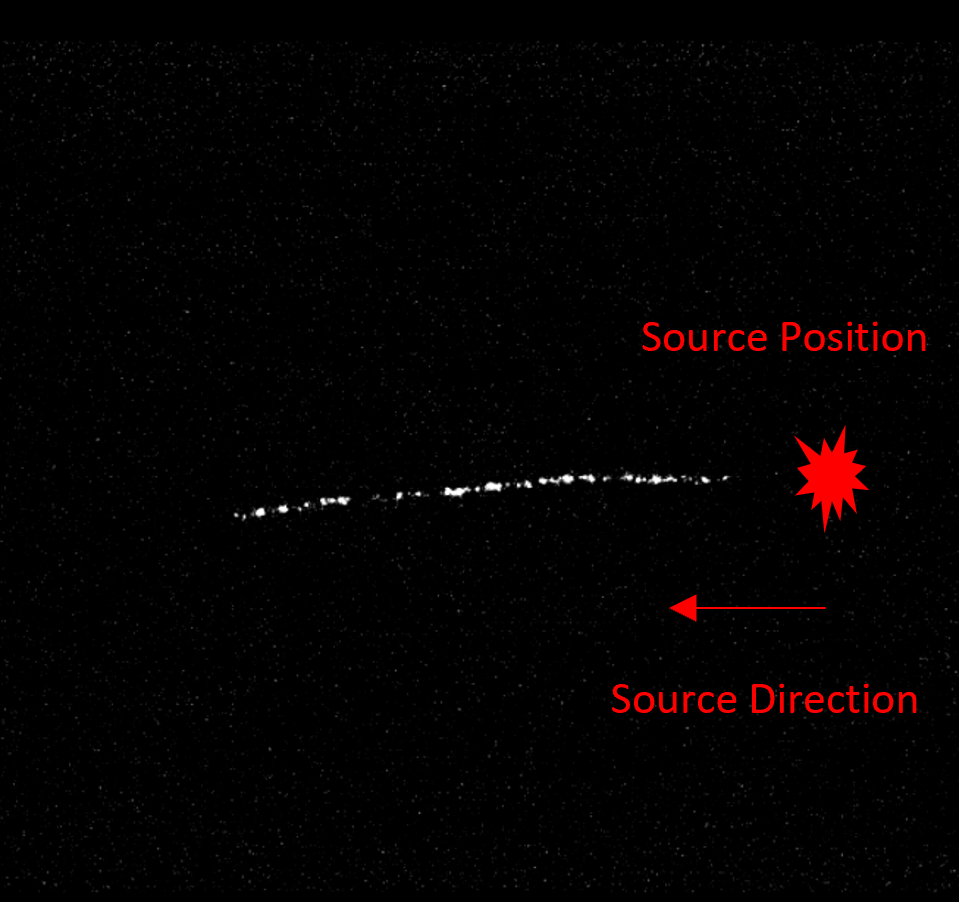}
        \caption{Contained electron track as seen by the camera. The source position and direction is highlighted.}
        \label{fig:image1}
    \end{minipage}
    \hfill
    \begin{minipage}[t]{0.3\textwidth}
        \centering
        \includegraphics[width=\textwidth]{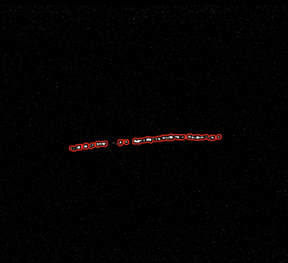}
        \caption{First step on the reconstruction algorithm, clusters of pixels over the threshold are grouped.}
        \label{fig:image2}
    \end{minipage}
    \hfill
    \begin{minipage}[t]{0.28\textwidth}
        \centering
        \includegraphics[width=\textwidth]{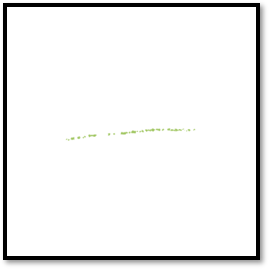}
        \caption{Directional part of the reconstruction algorithm merges the clusters that come from the same track.}
        \label{fig:image3}
    \end{minipage}
    \caption{Reconstruction algorithm workflow.}
    \label{fig:sidebyside}
\end{figure}

\begin{figure}[ht]
     \centering
     \includegraphics[width=\textwidth]{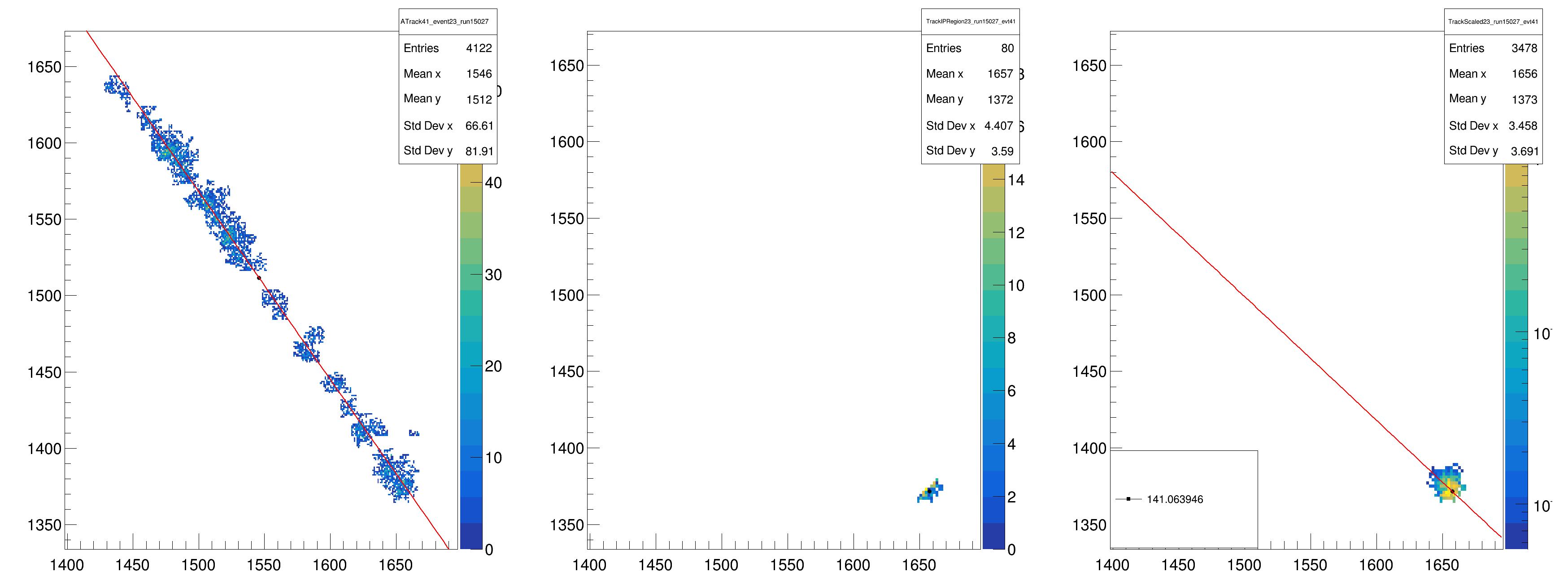}
     \caption{Directional algorithm on a fully contained track. Left: the original track with the barycenter and the principal axis. Center: the Interaction Point found by the algorithm. Right: selected and weighted points for direction computation with the estimated real direction.}
     \label{fig:dir}
\end{figure}

\begin{figure}[ht!]
    \centering
    \begin{minipage}[t]{0.45\textwidth}
        \centering
        \includegraphics[width=\textwidth]{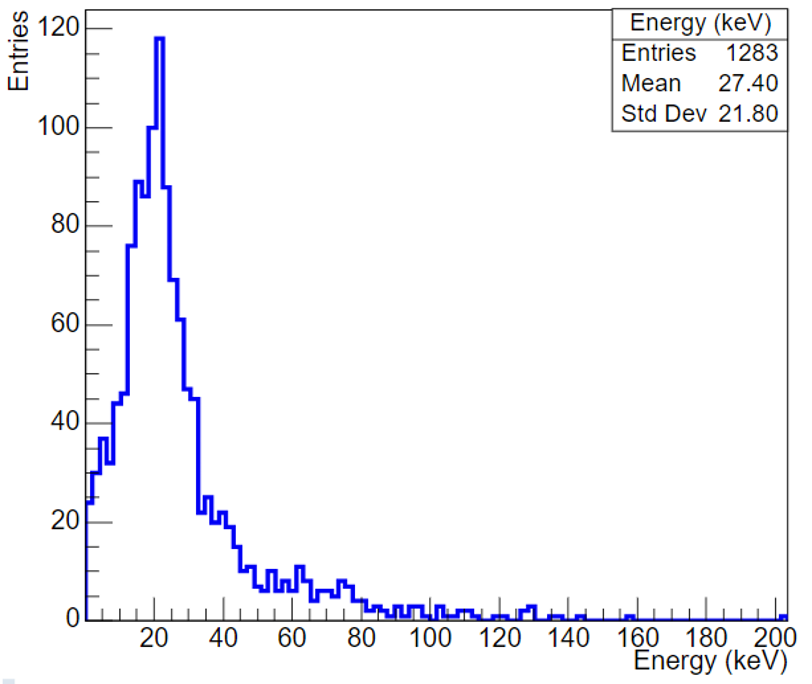}
        \caption{Calibrated energy spectrum of fully-contained tracks with correctly reconstructed IP for the data acquired with the $ He/CF_4$ 60/40 gas mixture.}
        \label{fig:spectrum}
    \end{minipage}
    \hfill
    \begin{minipage}[t]{0.45\textwidth}
        \centering
        \includegraphics[width=\textwidth]{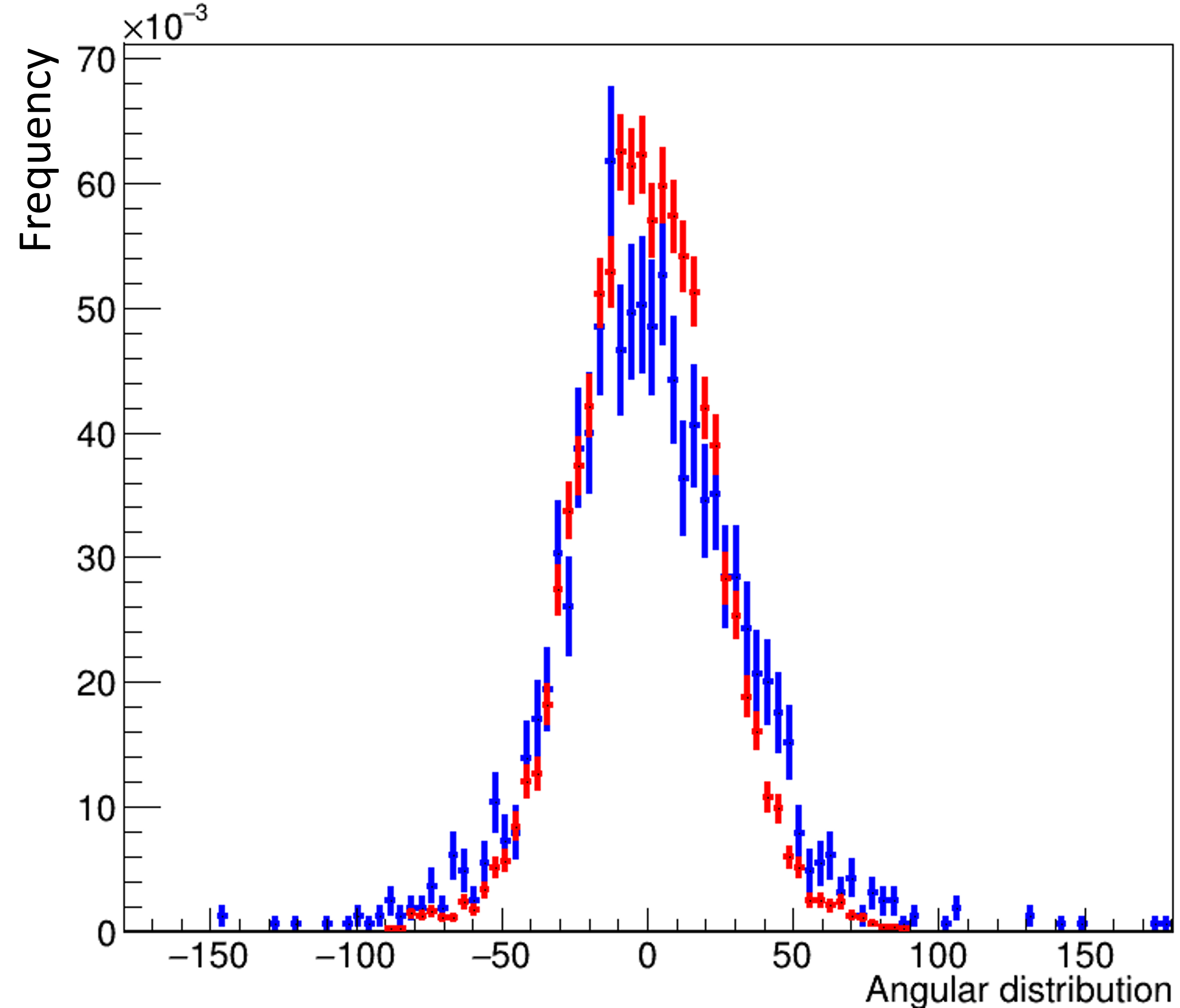}
        \caption{Intrinsic angular spread (RMS $\simeq 24^\circ$, red) from Geant4 vs. measured spread (RMS $\simeq 32^\circ$, blue) for the $ He/CF_4$ 60/40 gas mixture.}
        \label{fig:comparison}
    \end{minipage}
\end{figure}

Low-energy electrons emitted by the collimated $^{90}Sr$ source, primarily in a direction parallel to the GEM plane, were measured to evaluate the performance of the experimental approach in determining their initial direction. An example of a recorded track from an sCMOS image of a low-energy electron emitted by the $^{90}Sr$ source is displayed in Figure \ref{fig:image1}. The sCMOS images are analyzed and the tracks are reconstructed using the Directional iDBSCAN algorithm developed by the CYGNO collaboration \cite{dbscan_2,db_scan1}. The algorithm groups pixel clusters that exceed a threshold using DBSCAN (Figure \ref{fig:image2}), taking into account directional information to merge clusters originating from the same track (Figure \ref{fig:image3}). 
A dedicated algorithm based on Principal Component Analysis and adapted from the one used by the IXPE collaboration \cite{BALDINI2021102628} was optimized for the recorded sCMOS images to reconstruct the impact point and the direction of the measured tracks \cite{torelli2024feasibilitydirectionalsolarneutrino}. The steps of the algorithm are shown in Figure \ref{fig:dir}.
After all the tracks were reconstructed and their directions estimated, selection cuts were applied to clean the sample for further analysis. Quality selection cuts retain events with interaction points (IP) within a radius of $\rm 5~mm$ around the source position (to suppress natural radioactive background contamination) and reject tracks with hits within 5 mm of the sensitive volume border (to ensure containment). Figure \ref{fig:spectrum} displays the energy spectrum of the electron tracks resulting from this selection for the data acquired with the gas mixture $ He/CF_4$ 60/40.

As discussed previously, a simulation of the intrinsic angular spread of the electrons, including the collimator aperture, was performed to estimate their intrinsic angular spread. Figure \ref{fig:detGeant} shows a simulation schematic, in which the sensitive cylindrical volume is modeled together with the field cage rings made of PMMA with an internal silver electrode. We considered the electron tracks only within the sensitive volume. In fact, the same selection cuts used in the data analysis (full containment) are also applied to the simulated events. The initial direction of each electron track is determined by fitting a straight line to the projection of the first four hits on the GEM plane within the sensitive volume. The intrinsic spread obtained in this way, with an RMS of $\rm 23.6 \pm 0.2^\circ$, is shown in Figure \ref{fig:comparison}, alongside the measured spread of $\rm 32.3 \pm 0.8^\circ$.

\begin{figure}[t!]
    \centering
    \begin{minipage}[t]{0.45\textwidth}
        \centering
        \includegraphics[width=\textwidth]{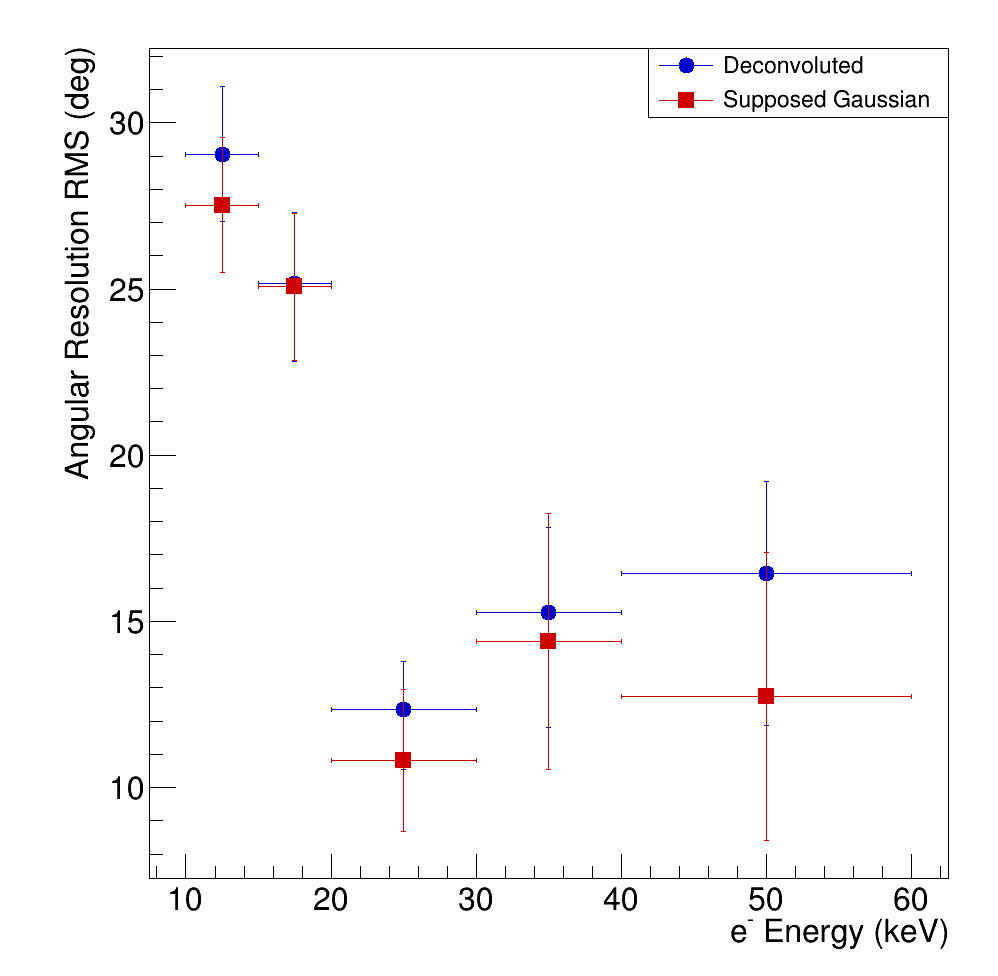}
        \caption{Measured angular resolution as a function of the calibrated electron energy. Blue: resolution obtained by deconvolution. Red: resolution obtained by quadratic subtraction of the measured and intrinsic spreads.}
        \label{fig:angularres}
    \end{minipage}
    \hfill
    \begin{minipage}[t]{0.45\textwidth}
        \centering
        \includegraphics[width=\textwidth]{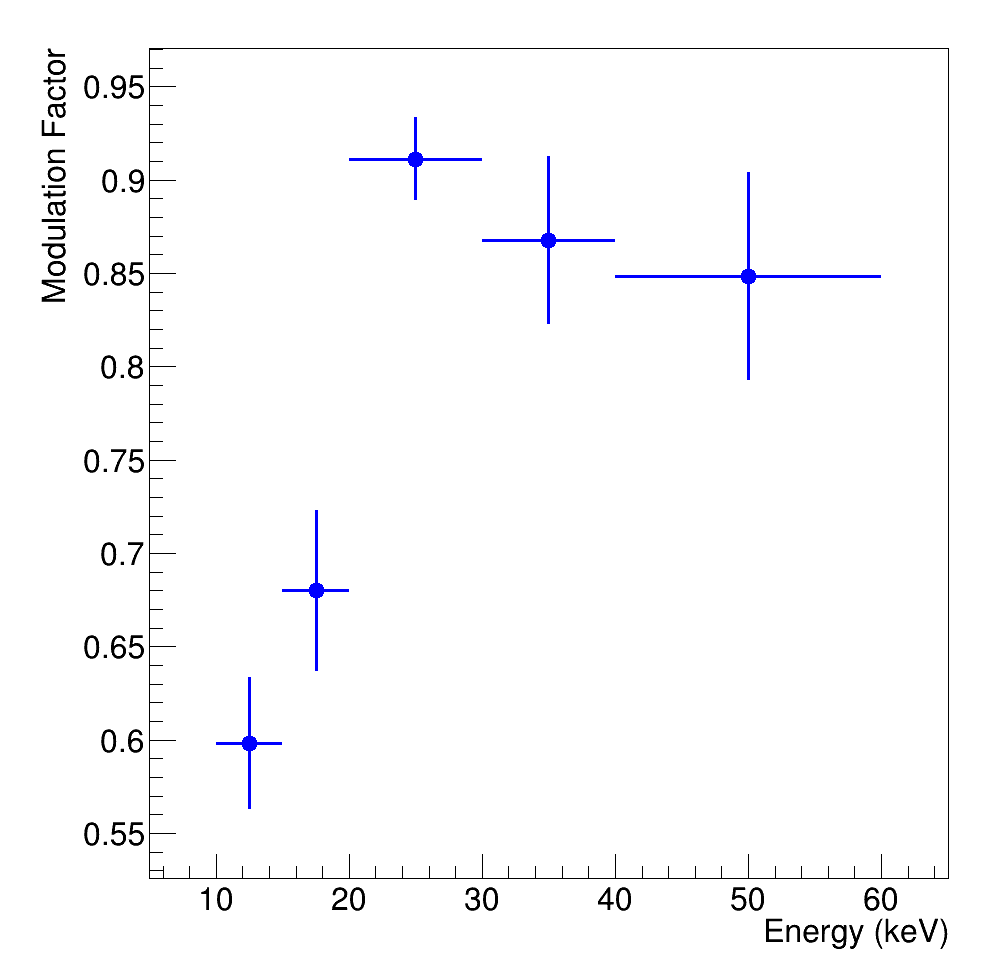}
        \caption{Expected modulation factor as a function of the X-ray energy for an optical readout polarimeter.}
        \label{fig:modfactor}
    \end{minipage}
\end{figure}

The measured angular distribution is then deconvolved from the intrinsic angular spread obtained from the GEANT4 simulation to extract the actual detector angular resolution. This deconvolution is performed using the Richardson–Lucy iterative method \cite{Richardson,lucy}. The RMS of the deconvolved distribution corresponds to the angular resolution of the detector, which is reported in Figure \ref{fig:angularres} for different energy bins. Additionally, Figure \ref{fig:angularres} shows the resolution obtained from a simple quadratic subtraction of the RMS values:
\[
RMS_{detector} = \sqrt{RMS_{measured}^2 - RMS_{intrinsic}^2},
\]
which assumes that both distributions are perfectly Gaussian. The close agreement between the two approaches confirms the validity of the deconvolution procedure.

With the reasonable assumption that similar performance can be achieved for photoelectrons emitted from the absorption of polarized X-rays, the measured angular resolution can be treated as a Gaussian response applied to $\cos^2$-modulated electrons. The expected distribution is expressed as:
$$
(G \ast \cos^2)(x) = \frac{1}{2} \left(1 + e^{-2\sigma^2} \cos(2x)\right).
$$
The modulation factor, $\mu$, can then be calculated using the formula:
$$
\mu = \frac{\frac{N^{\text{max}}}{100\%} - \frac{N^{\text{min}}}{100\%}}{\frac{N^{\text{max}}}{100\%} + \frac{N^{\text{min}}}{100\%}}.
$$
Figure \ref{fig:modfactor} shows the results.
These preliminary results indicate that a small TPC prototype, based on the CYGNO experimental approach with a cylindrical sensitive volume of $\rm 5~cm$ in height and $\rm 3.7~cm$ in radius, operating with an He/$CF_4$ (60/40) gas mixture, may achieve a modulation factor greater than 0.6 for X-rays above $\rm 10~keV$ and greater than 0.8 for X-rays above $\rm 20~keV$.

\subsection{Detector polarimetric sensivity estimation}\label{sec:results}
The Minimum Detectable Polarization (MDP) is a critical parameter for assessing the performance of a potential polarimeter. In the absence of background noise, the MDP can be expressed as \cite{figureof}:
$$
MDP = \frac{4.29}{\mu \sqrt{N}},
$$
where \(N\) is the number of photons detected from the source.
An effective polarimeter requires both a high modulation factor \(\mu\) and a high quantum efficiency \(\epsilon\), as these reduce the exposure time needed for a given source. Therefore, a useful figure of merit for characterizing a polarimeter is \(\mu\sqrt{\epsilon}\).

\begin{figure}[t!]
    \centering
    \begin{minipage}[t]{0.45\textwidth}
        \centering
        \includegraphics[width=\textwidth]{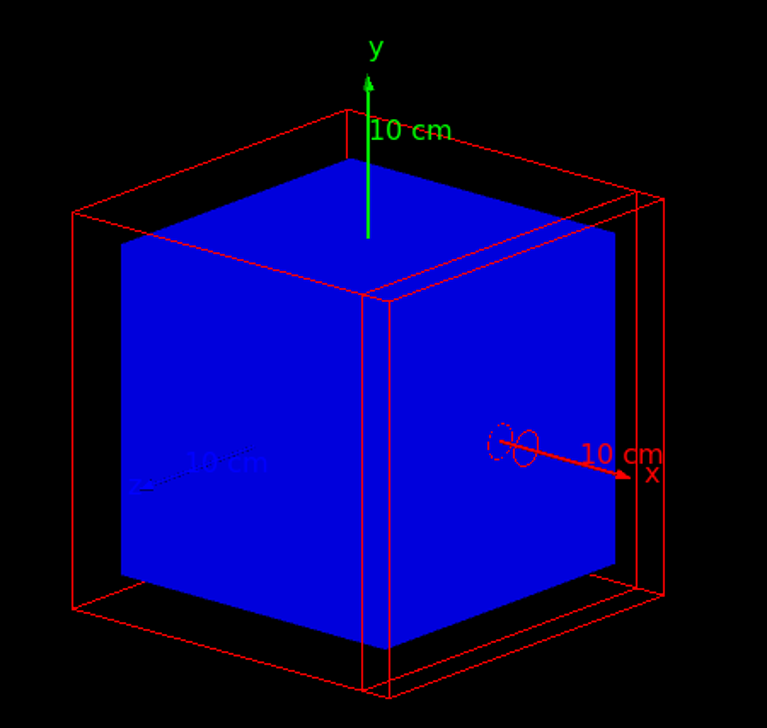}
        \caption{Simulated geometry of a simple collimated polarimeter.}
        \label{fig:simeff}
    \end{minipage}
    \hfill
    \begin{minipage}[t]{0.45\textwidth}
        \centering
        \includegraphics[width=\textwidth]{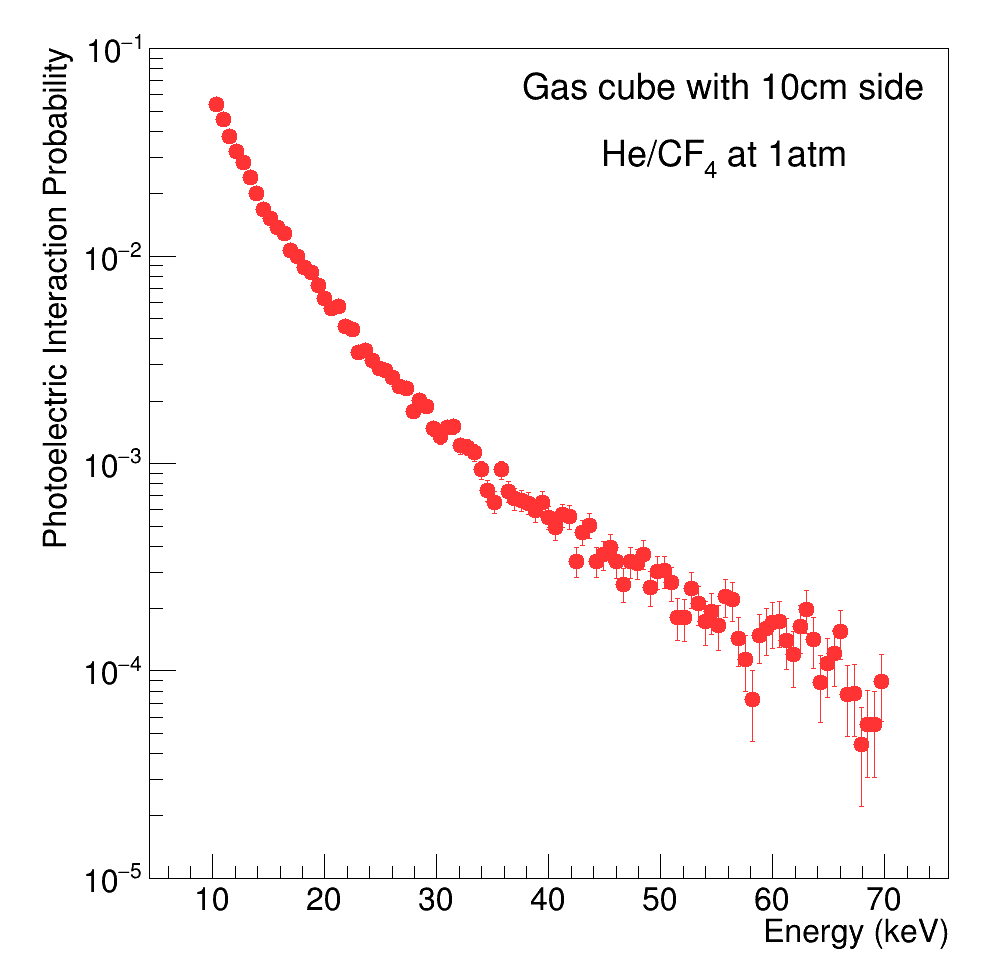}
        \caption{Simulated photoelectric efficiency.}
        \label{fig:efficeicny}
    \end{minipage}
\end{figure}

We performed a Geant4 simulation to estimate the efficiency of a collimated polarimeter based on the results presented in Sec \ref{sec:analysis}, with the goal of developing a compact instrument with a wide field of view. In the simulation, we modeled a gas cube with $\rm 10~cm$ sides, surrounded by $\rm 2~cm$ aluminum shielding. The entrance window, placed in the center of one side of the shielding, has a diameter of $\rm 10.6~mm$, corresponding to a potential maximum field of view of $\rm 27^\circ$. Although a polarimeter wthin a realistic experiment could include would include a Micro Channel Plate (MCP) to select only photons incident perpendicularly on the entrance window, the MCP was omitted for simplicity in the simulation. Figure \ref{fig:simeff} illustrates the simulated geometry.

Photons with energies between $\rm 10~keV$ and $\rm 60~keV$ were simulated to enter through the entrance window or a larger delimiter of a wide field. Secondary particles were tracked, and two conditions had to be satisfied for a photon to be counted as detected. First, the total deposited energy from the photon and its secondaries had to match the initial photon energy within $\rm 200~eV$, to account for binding energies. Second, the initial photon had to interact only once in the sensitive volume, ensuring that it underwent a single photoelectric interaction. This assumption implicitly includes the detector’s capability to reject events where the photon undergoes Compton scattering in the gas cell and then escapes, for example, by means of an active anti-coincidence system.

Figure \ref{fig:efficeicny} shows the simulated efficiency for this configuration. The efficiency is relatively low for such a detector design, mainly due to the low density and atomic number of the CYGNO gas mixture, which was optimized for dark matter searches. To explore alternatives, we tested different gas mixtures by varying density and composition. Increasing the CF\(_4\) content improves the efficiency approximately linearly, while replacing He with a higher-$Z$ element such as Ar further enhances efficiency, owing to the $Z^5$ dependence of the photoelectric cross section.

\begin{figure}[ht!]
     \centering
     \begin{minipage}[t]{0.45\textwidth}
         \centering
         \includegraphics[width=\textwidth]{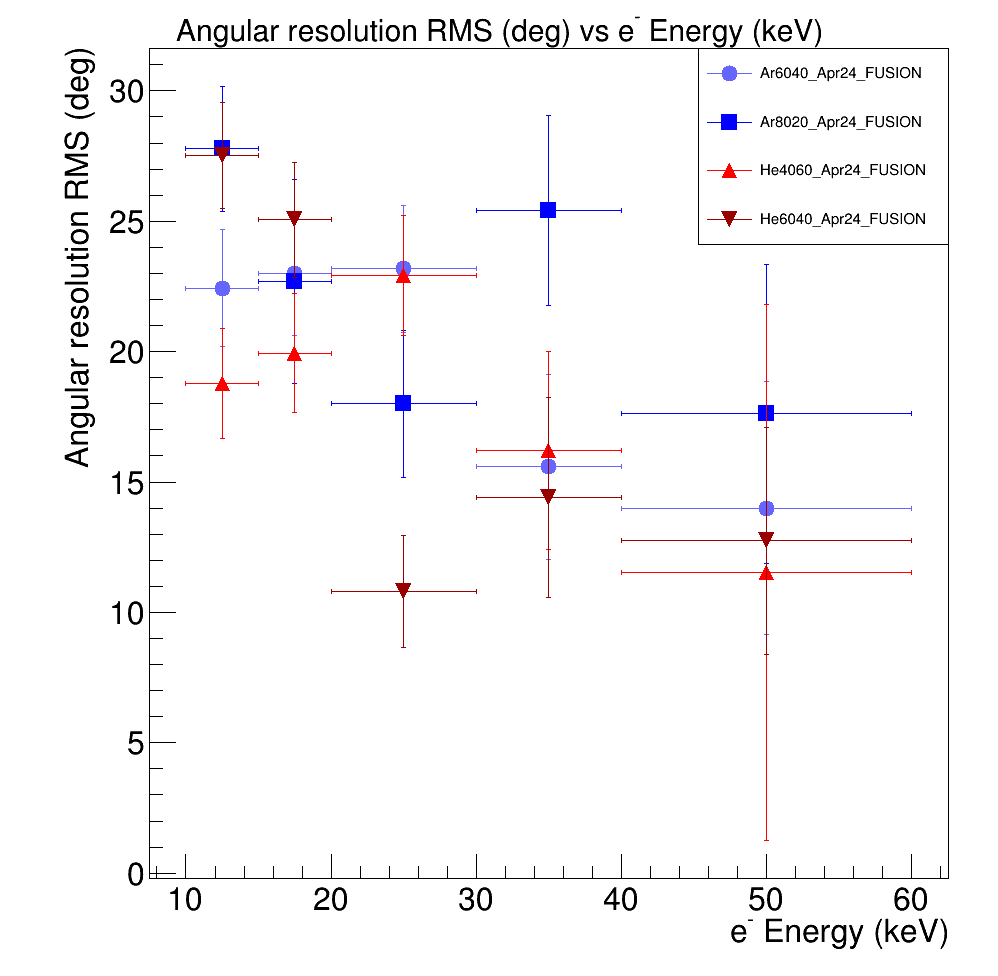}
         \caption{Measured angular resolutions for different gas mixtures.}
         \label{fig:angresgases}
     \end{minipage}
     \hfill
     \begin{minipage}[t]{0.45\textwidth}
         \centering
         \includegraphics[width=\textwidth]{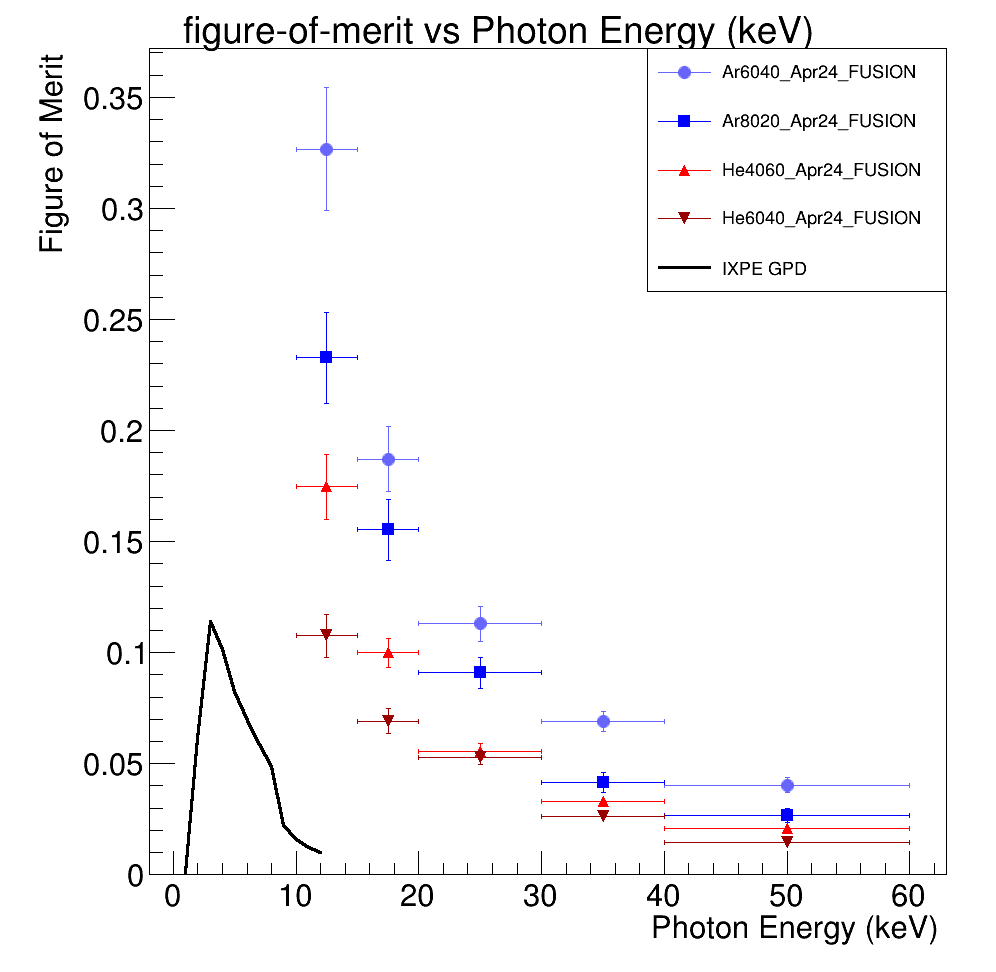}
         \caption{Expected figure of merit $\mu\sqrt{\epsilon}$ for the tested gas mixtures. The IXPE GPD figure of merit is also added for comparison.}
         \label{fig:figuregases}
     \end{minipage}
 \end{figure}

The data collection and analysis procedure described above was then repeated for several gas mixtures at ambient pressure: He/\(CF_4\) (40/60), Ar/\(CF_4\) (60/40), and Ar/\(CF_4\) (80/20). Figure \ref{fig:angresgases} shows the angular resolutions obtained for the four gas mixtures. The angular resolution does not change significantly between mixtures, as it is primarily determined by the detector settings and analysis method, which are common to all. The impact of the gas mixture is more evident in Figure \ref{fig:figuregases}, where the higher efficiency of Ar-based mixtures improves the figure of merit by almost a factor of four compared to the CYGNO gas mixture. The figure of merit for IXPE GPD is also added to the plot for comparison, data are extracted from ixpeobssimm \cite{BALDINI2022101194}. This translates into a substantial reduction in observation time, highlighting the advantage of Argon for enhancing detection performance.

\section{Conclusion and future perspectives}
We presented the first results from a triple-GEM TPC with optical readout, inspired by the CYGNO directional Dark Matter search approach, applied to hard X-ray polarimetry. Using collimated $^{90}$Sr electrons, we demonstrated angular resolutions better than $30^\circ$ above 10\,keV and better than $20^\circ$ for 20–60 keV electrons. Folding these resolutions with a $\cos^2$ response yields inferred modulation factors $\mu \gtrsim 0.6$ above 10 keV and $\mu \gtrsim 0.8$ for 20–60 keV. Further improvements in the proposed experimental approach performances are expected from optimization of gas pressure/density/composition (e.g., Ar-rich mixtures), as well as hardware and reconstruction improvements (3D timing via PMT, next-generation sCMOS, dedicated tracking). In addition to the results reported here, first measurements with fully polarized beams have already been performed at the INAF calibration facility in Rome \cite{Muleri2022CalibrationFacility}, providing actual indications of the achievable realistic performance as a polarimeter, and a dedicated paper is in preparation.

These results strongly support the feasibility of a CYGNO-like optical TPC as a wide-FoV hard X-ray polarimeter. Such a detector could enable polarization measurements of rapid and unpredictable high-energy transients, such as gamma-ray bursts and solar flares \cite{Toma2009ApJ}, magnetar giant flares \cite{Mereghetti2008}, tidal disruption events \cite{Komossa2015}, or multi-messenger sources including binary neutron star mergers \cite{Abbott2017GW170817}. By opening sensitivity to such short-lived and rare phenomena, this technique would extend the reach of X-ray polarimetry into domains inaccessible to current narrow-field missions like IXPE, and provide unique synergies with future observatories across the electromagnetic and multi-messenger spectrum.

\acknowledgments

This work is supported by the Italian Ministry of Education, University and Research (MIUR) through the PRIN project “HypeX: High Yield Polarimetry Experiment in X-rays” (Prot.\ 2020MZ884C\_001).


\bibliographystyle{JHEP}
\bibliography{biblio.bib}

@INPROCEEDINGS{Austin1994,
       author = {{Austin}, Robert A. and {Minamitani}, Takahisa and {Ramsey}, Brian D.},
        title = "{Development of a hard x-ray imaging polarimeter}",
    booktitle = {X-Ray and Ultraviolet Polarimetry},
         year = 1994,
       editor = {{Fineschi}, Silvano},
       series = {Society of Photo-Optical Instrumentation Engineers (SPIE) Conference Series},
       volume = {2010},
        month = feb,
        pages = {118-125},
          doi = {10.1117/12.168571},
       adsurl = {https://ui.adsabs.harvard.edu/abs/1994SPIE.2010..118A}
}

@article{10.1117/1.JATIS.8.2.026002,
author = {{Weisskopf}, Martin and others},
title = {{Imaging X-ray Polarimetry Explorer: prelaunch}},
volume = {8},
journal = {Journal of Astronomical Telescopes, Instruments, and Systems},
number = {2},
publisher = {SPIE},
pages = {026002},
year = {2022},
doi = {10.1117/1.JATIS.8.2.026002},
URL = {https://doi.org/10.1117/1.JATIS.8.2.026002}
}

@ARTICLE{2021Soffitta,
       author = {{Soffitta}, Paolo and {Baldini}, Luca and {Bellazzini}, Ronaldo and {Costa}, Enrico and {Latronico}, Luca and {Muleri}, Fabio and {Del Monte}, Ettore and {Fabiani}, Sergio and {Minuti}, Massimo and {Pinchera}, Michele and {Sgro'}, Carmelo and {Spandre}, Gloria and {Trois}, Alessio and {Amici}, Fabrizio and {Andersson}, Hans and {Attina'}, Primo and {Bachetti}, Matteo and {Barbanera}, Mattia and {Borotto}, Fabio and {Brez}, Alessandro and {Brienza}, Daniele and {Caporale}, Ciro and {Cardelli}, Claudia and {Carpentiero}, Rita and {Castellano}, Simone and {Castronuovo}, Marco and {Cavalli}, Luca and {Cavazzuti}, Elisabetta and {Ceccanti}, Marco and {Centrone}, Mauro and {Ciprini}, Stefano and {Citraro}, Saverio and {D'Amico}, Fabio and {D'Alba}, Elisa and {Di Cosimo}, Sergio and {Di Lalla}, Niccolo' and {Di Marco}, Alessandro and {Di Persio}, Giuseppe and {Donnarumma}, Immacolata and {Evangelista}, Yuri and {Ferrazzoli}, Riccardo and {Hayato}, Asami and {Kitaguchi}, Takao and {La Monaca}, Fabio and {Lefevre}, Carlo and {Loffredo}, Pasqualino and {Lorenzi}, Paolo and {Lucchesi}, Leonardo and {Magazzu}, Carlo and {Maldera}, Simone and {Manfreda}, Alberto and {Mangraviti}, Elio and {Marengo}, Marco and {Matt}, Giorgio and {Mereu}, Paolo and {Morbidini}, Alfredo and {Mosti}, Federico and {Nakano}, Toshio and {Nasimi}, Hikmat and {Negri}, Barbara and {Nenonen}, Seppo and {Nuti}, Alessio and {Orsini}, Leonardo and {Perri}, Matteo and {Pesce-Rollins}, Melissa and {Piazzolla}, Raffaele and {Pilia}, Maura and {Profeti}, Alessandro and {Puccetti}, Simonetta and {Rankin}, John and {Ratheesh}, Ajay and {Rubini}, Alda and {Santoli}, Francesco and {Sarra}, Paolo and {Scalise}, Emanuele and {Sciortino}, Andrea and {Tamagawa}, Toru and {Tardiola}, Marcello and {Tobia}, Antonino and {Vimercati}, Marco and {Xie}, Fei},
        title = "{The Instrument of the Imaging X-Ray Polarimetry Explorer}",
      journal = {Astronomical Journal},
         year = 2021,
        month = nov,
       volume = {162},
       number = {5},
          eid = {208},
        pages = {208},
          doi = {10.3847/1538-3881/ac19b0},
archivePrefix = {arXiv},
       eprint = {2108.00284},
 primaryClass = {astro-ph.IM},
       adsurl = {https://ui.adsabs.harvard.edu/abs/2021AJ....162..208S}
}

@article{amaro2022cygno,
  title={The CYGNO experiment},
  author={Amaro, Fernando Domingues and Baracchini, Elisabetta and Benussi, Luigi and Bianco, Stefano and Capoccia, Cesidio and Caponero, Michele and Cardoso, Danilo Santos and Cavoto, Gianluca and Cortez, Andr{\'e} and Costa, Igor Abritta and others},
  journal={Instruments},
  volume={6},
  number={1},
  pages={6},
  year={2022},
  publisher={MDPI}
}

@article{commissionig,
  title={A 50 l CYGNO prototype overground characterization},
  author={Amaro, F.D. and Antonietti, R. and Baracchini, E. and others},
  journal={Eur. Phys. J. C},
  volume = {83},
  pages = {946},
  year={2023},
  doi = {10.1140/epjc/s10052-023-11988-9},
  URL={https://doi.org/10.1140/epjc/s10052-023-11988-9}
}

@article{db_scan1,
doi = {10.1088/1361-6501/acf402},
url = {https://dx.doi.org/10.1088/1361-6501/acf402},
year = {2023},
month = {sep},
publisher = {IOP Publishing},
volume = {34},
number = {12},
pages = {125024},
author = {F D Amaro and R Antonietti and E Baracchini  and otherss},
title = {Directional iDBSCAN to detect cosmic-ray tracks for the CYGNO experiment},
journal = {Measurement Science and Technology}
}

@article{dbscan_2,
doi = {10.1088/1748-0221/15/12/T12003},
url = {https://dx.doi.org/10.1088/1748-0221/15/12/T12003},
year = {2020},
month = {dec},
publisher = {},
volume = {15},
number = {12},
pages = {T12003},
author = {E. Baracchini  and others},
title = {A density-based clustering algorithm for the CYGNO data analysis},
journal = {Journal of Instrumentation}
}

@article{Sauli:1997qp,
    author = "Sauli, F.",
    title = "{GEM: A new concept for electron amplification in gas detectors}",
    doi = "10.1016/S0168-9002(96)01172-2",
    journal = "Nucl. Instrum. Meth. A",
    volume = "386",
    pages = "531--534",
    year = "1997"
}

@article{geant,
title = {Geant4—a simulation toolkit},
journal = {Nuclear Instruments and Methods in Physics Research Section A: Accelerators, Spectrometers, Detectors and Associated Equipment},
volume = {506},
number = {3},
pages = {250-303},
year = {2003},
issn = {0168-9002},
doi = {https://doi.org/10.1016/S0168-9002(03)01368-8},
url = {https://www.sciencedirect.com/science/article/pii/S0168900203013688},
author = {S. Agostinelli  and others}
}

@inproceedings{figureof,
author = {Martin C. Weisskopf and Ronald F. Elsner and Stephen L. O'Dell},
title = {{On understanding the figures of merit for detection and measurement of x-ray polarization}},
volume = {7732},
booktitle = {Space Telescopes and Instrumentation 2010: Ultraviolet to Gamma Ray},
editor = {Monique Arnaud and Stephen S. Murray and Tadayuki Takahashi},
organization = {International Society for Optics and Photonics},
publisher = {SPIE},
pages = {77320E},
year = {2010},
doi = {10.1117/12.857357},
URL = {https://doi.org/10.1117/12.857357}
}

@article{BALDINI2021102628,
title = {Design, construction, and test of the Gas Pixel Detectors for the IXPE mission},
journal = {Astroparticle Physics},
volume = {133},
pages = {102628},
year = {2021},
issn = {0927-6505},
doi = {https://doi.org/10.1016/j.astropartphys.2021.102628},
url = {https://www.sciencedirect.com/science/article/pii/S0927650521000670},
author = {L. Baldini  and others}
}

@ARTICLE{lucy,
  author = {Lucy, L.~B.},
  title = {An iterative technique for the rectification of observed distributions},
  journal = {The Astronomical Journal},
  year = {1974},
  month = jun,
  volume = {79},
  pages = {745},
  doi = {10.1086/111605},
  adsurl = {https://ui.adsabs.harvard.edu/abs/1974AJ.....79..745L}
}

@article{Richardson,
  author = {William Hadley Richardson},
  journal = {J. Opt. Soc. Am.},
  number = {1},
  pages = {55--59},
  publisher = {Optica Publishing Group},
  title = {Bayesian-Based Iterative Method of Image Restoration},
  volume = {62},
  month = {Jan},
  year = {1972},
  url = {https://opg.optica.org/abstract.cfm?URI=josa-62-1-55},
  doi = {10.1364/JOSA.62.000055}
}

@article{FRAGA200388,
title = {The GEM scintillation in He–CF4, Ar–CF4, Ar–TEA and Xe–TEA mixtures},
journal = {Nuclear Instruments and Methods in Physics Research Section A: Accelerators, Spectrometers, Detectors and Associated Equipment},
volume = {504},
number = {1},
pages = {88-92},
year = {2003},
note = {Proceedings of the 3rd International Conference on New Developments in Photodetection},
issn = {0168-9002},
doi = {https://doi.org/10.1016/S0168-9002(03)00758-7},
url = {https://www.sciencedirect.com/science/article/pii/S0168900203007587},
author = {M.M.F.R. Fraga and F.A.F. Fraga and S.T.G. Fetal and L.M.S. Margato and R.Ferreira Marques and A.J.P.L. Policarpo}
}

@misc{torelli2024feasibilitydirectionalsolarneutrino,
      title={Feasibility of a directional solar neutrino measurement with the CYGNO/INITIUM experiment}, 
      author={Samuele Torelli},
      year={2024},
      eprint={2408.03760},
      archivePrefix={arXiv},
      primaryClass={physics.ins-det},
      url={https://arxiv.org/abs/2408.03760}, 
}

@article{Costa2001Nature,
  author       = {Costa, E. and Soffitta, P. and Bellazzini, R. and Brez, A. and Lumb, N. and Spandre, G.},
  title        = {An efficient photoelectric X-ray polarimeter for the study of black holes and neutron stars},
  journal      = {Nature},
  year         = {2001},
  volume       = {411},
  pages        = {662--665},
  doi          = {10.1038/35079508}
}

@ARTICLE{Sakurai2004,
       author = {{Sakurai}, H. and {Tokanai}, F. and {Gunji}, S. and {Motegi}, S. and {Toyokawa}, H. and {Suzuki}, M. and {Hirota}, K. and {Kishimoto}, S.},
        title = "{Measurement of X-ray polarization using optical imaging detector with capillary plate}",
      journal = {Nuclear Instruments and Methods in Physics Research A},
         year = 2004,
        month = jun,
       volume = {525},
       number = {1-2},
        pages = {6-11},
          doi = {10.1016/j.nima.2004.03.132},
       adsurl = {https://ui.adsabs.harvard.edu/abs/2004NIMPA.525....6S}
}

@inproceedings{Weisskopf2010SPIE,
  author       = {Weisskopf, M. C. and Elsner, R. F. and O’Dell, S. L.},
  title        = {On understanding the figures of merit for detection and measurement of X-ray polarization},
  booktitle    = {Proc. SPIE 7732, Space Telescopes and Instrumentation 2010: Ultraviolet to Gamma Ray},
  year         = {2010},
  volume       = {7732},
  pages        = {77320E},
  doi          = {10.1117/12.857357}
}

@article{Toma2009ApJ,
  author       = {Toma, K. and Sakamoto, T. and Zhang, B. and Hill, J. E. and McTiernan, J. and Norris, J. P. and Gehrels, N.},
  title        = {Statistical Properties of Gamma-Ray Burst Polarization},
  journal      = {Astrophysical Journal},
  year         = {2009},
  volume       = {698},
  number       = {2},
  pages        = {1042--1053},
  doi          = {10.1088/0004-637X/698/2/1042}
}

@article{Kislat2015APh,
  author       = {Kislat, Fabian and Beilicke, Matthias and Guo, Qian and Krawczynski, Henric},
  title        = {Analyzing the data from X-ray polarimeters with Stokes parameters},
  journal      = {Astroparticle Physics},
  year         = {2015},
  volume       = {68},
  pages        = {45--51},
  doi          = {10.1016/j.astropartphys.2015.02.007}
}

@article{Muleri2022CalibrationFacility,
  author       = {Muleri, Fabio and Piazzolla, Raffaele and Di Marco, Alessandro and Fabiani, Sergio and La Monaca, Fabio and et al.},
  title        = {The IXPE Instrument Calibration Equipment},
  journal      = {Nuclear Instruments and Methods in Physics Research A},
  year         = {2022},
  doi          = {DOI_from_original_source}
}

@article{Mereghetti2008,
  author       = {Mereghetti, Sandro},
  title        = {The strongest cosmic magnets: soft gamma-ray repeaters and anomalous X-ray pulsars},
  journal      = {Astronomy and Astrophysics Review},
  year         = {2008},
  volume       = {15},
  pages        = {225--287},
  doi          = {10.1007/s00159-008-0011-z}
}

@article{Komossa2015,
  author       = {Komossa, S.},
  title        = {Tidal disruption of stars by supermassive black holes: Status of observations},
  journal      = {Journal of High Energy Astrophysics},
  year         = {2015},
  volume       = {7},
  pages        = {148--157},
  doi          = {10.1016/j.jheap.2015.04.006}
}

@article{Abbott2017GW170817,
  author       = {Abbott, B. P. and Abbott, R. and Abbott, T. D. and et al. (LIGO Scientific Collaboration and Virgo Collaboration)},
  title        = {GW170817: Observation of Gravitational Waves from a Binary Neutron Star Inspiral},
  journal      = {Physical Review Letters},
  year         = {2017},
  volume       = {119},
  pages        = {161101},
  doi          = {10.1103/PhysRevLett.119.161101}
}

@article{CORRADI200796,
	title = {A novel High-Voltage System for a triple GEM detector},
	journal = {Nuclear Instruments and Methods in Physics Research Section A: Accelerators, Spectrometers, Detectors and Associated Equipment},
	volume = {572},
	number = {1},
	pages = {96-97},
	year = {2007},
	note = {Frontier Detectors for Frontier Physics},
	issn = {0168-9002},
	doi = {https://doi.org/10.1016/j.nima.2006.10.166},
	url = {https://www.sciencedirect.com/science/article/pii/S0168900206020146},
	author = {G. Corradi and F. Murtas and D. Tagnani}
}

@article{BALDINI2022101194,
title = {ixpeobssim: A simulation and analysis framework for the imaging X-ray polarimetry explorer},
journal = {SoftwareX},
volume = {19},
pages = {101194},
year = {2022},
issn = {2352-7110},
doi = {https://doi.org/10.1016/j.softx.2022.101194},
url = {https://www.sciencedirect.com/science/article/pii/S2352711022001169},
author = {Luca Baldini and Niccolò Bucciantini and Niccolò Di Lalla and Steven Ehlert and Alberto Manfreda and Michela Negro and Nicola Omodei and Melissa Pesce-Rollins and Carmelo Sgrò and Stefano Silvestri}
}

@inproceedings{Soffitta2012,
  author    = {Soffitta, Paolo and Campana, Riccardo and Costa, Enrico and Fabiani, Sergio and Muleri, Fabio and Rubini, Alda and Bellazzini, Ronaldo and Brez, Alessandro and Minuti, Massimo and Pinchera, Michele and Spandre, Gloria},
  title     = {The Background of the Gas Pixel Detector and its impact on imaging X-ray polarimetry},
  booktitle = {Proceedings of SPIE},
  volume    = {8443},
  year      = {2012},
  doi       = {10.1117/12.925385}
}

@ARTICLE{DiMarco2023,
	author = {Di Marco, Alessandro and Soffitta, Paolo and Costa, Enrico and Ferrazzoli, Riccardo and La Monaca, Fabio and Rankin, John and Ratheesh, Ajay and Xie, Fei and Baldini, Luca and Del Monte, Ettore and Ehlert, Steven R. and Fabiani, Sergio and Kim, Dawoon E. and Muleri, Fabio and O’Dell, Stephen L. and Ramsey, Brian D. and Rubini, Alda and Sgrò, Carmelo and Silvestri, Stefano and Tennant, Allyn F. and Weisskopf, Martin C.},
	title = {Handling the Background in IXPE Polarimetric Data},
	year = {2023},
	journal = {Astronomical Journal},
	volume = {165},
	number = {4},
	doi = {10.3847/1538-3881/acba0f},
	url = {https://www.scopus.com/inward/record.uri?eid=2-s2.0-85149552505&doi=10.3847%2f1538-3881%2facba0f&partnerID=40&md5=256bc724b31b9197354e4cc7a4eb9ed8},
	type = {Article},
	publication_stage = {Final},
	source = {Scopus},
	note = {Cited by: 98; All Open Access, Gold Open Access, Green Open Access}
}

@article{directionality,
    author = {Torelli, Samuele and Fiorina, Davide},
    title = {Directionality of low energy electron recoils in the CYGNO experiment},
    journal = {Under internal review},
    year = {2026}
}


\end{document}